\documentclass[9pt,twocolumn,twoside]{opti_preprint}
\setboolean{shortarticle}{true}

\usepackage{soul}
\usepackage{xr-hyper}

\title{Super-resolution photoacoustic fluctuation imaging with multiple speckle illumination}

\author[1,2,*,$\dagger$]{Thomas Chaigne}
\author[1,2,$\dagger$]{J\'er\^ome Gateau}
\author[3]{Marc Allain}
\author[1,2]{Ori Katz}
\author[2]{Sylvain Gigan}
\author[3]{Anne Sentenac}
\author[1]{Emmanuel Bossy}

\affil[1]{ESPCI ParisTech, PSL Research University, CNRS UMR 7587, INSERM U979, Institut Langevin, 1 rue Jussieu, 75005, Paris, France}
\affil[2]{Laboratoire Kastler Brossel, Universit\'e Pierre et Marie Curie, Ecole Normale Sup\'erieure, Coll\`ege de France, CNRS UMR 8552, 24 rue Lhomond 75005 Paris, France}
\affil[3]{Institut Fresnel (CNRS UMR 7249), Universit\'e Aix-Marseille, Campus de St J\'er\^ome, 13013 Marseille, France,}

\affil[*]{Corresponding author: thomas.chaigne@espci.fr}
\affil[$\dagger$]{These authors contributed equally to this work}

\begin{abstract}
In deep tissue photoacoustic imaging, the spatial resolution is inherently limited by acoustic diffraction. Moreover, as the ultrasound attenuation increases with frequency, resolution is often traded-off for penetration depth. Here we report on super-resolution photoacoustic imaging by use of multiple speckle illumination. Specifically, we show that the analysis of second-order fluctuations of the photoacoustic images combined with image deconvolution enables resolving optically absorbing structures beyond the acoustic diffraction limit. A resolution increase of almost a factor 2 is demonstrated experimentally. Our method introduces a new framework that could potentially lead to deep tissue photoacoustic imaging with sub-acoustic resolution.
\end{abstract}

\begin{document}

\maketitle
\pagestyle{plain}
\ifthenelse{\boolean{shortarticle}}{\abscontent}{}

Light scattering prevents standard optical microscopes to obtain well-resolved images deep inside biological tissues. In the past twenty years, photoacoustic (PA) imaging has been developed to overcome this limitation, by imaging optical absorption deep inside strongly scattering tissue with the resolution of ultrasound~\cite{beard2011biomedical}. PA imaging relies on the unscattered ultrasonic waves emitted by absorbing structures under pulsed illumination via thermo-elastic stress generation. It therefore provides images at depth in tissue with a spatial resolution limited by acoustic diffraction. Ultimately, the ultrasound resolution for biological soft tissue is limited by the attenuation of ultrasound, which typically increases linearly with frequency. As a result, the depth-to-resolution ratio of PA imaging at depth is around 200 in practice~\cite{beard2011biomedical,wang2012photoacoustic}. As an illustration, axial resolution down to 5 µm and lateral resolution  down to 10 µm have been reached with high frequency acoustic detectors at depth up to 5 mm~\cite{omar2014ultrawideband}.  

In this letter, we demonstrate that the conventional acoustic-diffraction limit in PA imaging may be overcome by exploiting PA signal fluctuations, building on the super-resolution optical fluctuation imaging (SOFI) technique developed for fluorescence microscopy~\cite{dertinger2009fast}. SOFI is based on the idea that a higher-order statistical analysis of \textit{temporal} fluctuations caused by fluorescence blinking provides a way to resolve uncorrelated fluorophores within a same diffraction spot. In this work, we introduce multiple optical speckle illumination as a source of fluctuations for  super-resolution PA imaging, inspired by the principle introduced in optics with SOFI~\citep{dertinger2009fast} or from derived approaches using speckle illumination\citep{oh2013sub}. In PA imaging, multiple speckle illumination was initially introduced by our group as a mean to palliate limited-view or highpass-filtering artefacts~\cite{gateau2013improving}. Here, we demonstrate that a second-order analysis of optical speckle-induced PA fluctuations also provides super-resolved PA images beyond the acoustic diffraction limit.

In this work, we consider PA images reconstructed from a set of PA signals measured with an ultrasound array. A conventional backprojection algorithm is used to reconstruct the images, and it is assumed that the reconstructed PA quantity $A(\mathbf{r})$ may be written as a convolution:
\begin{equation}
\label{eq:Reconstruction}
A(\mathbf{r})=\left[\mu_{a}(\mathbf{r})\times I(\mathbf{r})\right]\ast h(\mathbf{r})
\end{equation} 
where $h$ is the PSF corresponding to the conventional PA imaging process,  $\mu_{a}$ the distribution of optical absorption and $I$ the optical intensity pattern (see Supplement 1, sec. 1.A,  for a detailed justification of Eq.~\ref{eq:Reconstruction}). Let us now consider that the region of interest is successively illuminated by many different speckle patterns $I_k(\textbf{r})$ with mean $\langle I(\textbf{r}) \rangle  = I_0$. The following expression for the mean PA image (estimated from averaging the PA images obtained with many realizations $I_k(\textbf{r})$ of the speckle illumination) 
\begin{equation}
\label{mean}
\langle A \rangle (\mathbf{r})  = I_0 \times \left[\mu_{a}(\mathbf{r}) \ast h(\mathbf{r})\right]
\end{equation}
shows that the resolution of the reconstructed image $\langle A \rangle (\mathbf{r})$ is dictated by the spatial frequency content of $h(\mathbf{r})$. Under the assumption that the optical speckle size is much smaller than that of $h(\mathbf{r})$, the variance image $\sigma^2[A](\mathbf{r})$ for uncorrelated speckles is given by (see Supplement 1, sec. 1.B) 
\begin{equation}
  \label{eq:var}
\sigma^2[A](\mathbf{r})  \propto \mu_{a} ^2(\mathbf{r}) \ast h^2(\mathbf{r})
\end{equation}
The variance image appears as the convolution of the squared object by the squared PSF, which has a higher frequency content than the PSF itself. As a result, the variance image is expected to have a higher resolution as compared to the mean image.  

\begin{figure}
\centerline{\includegraphics[width=1\columnwidth]{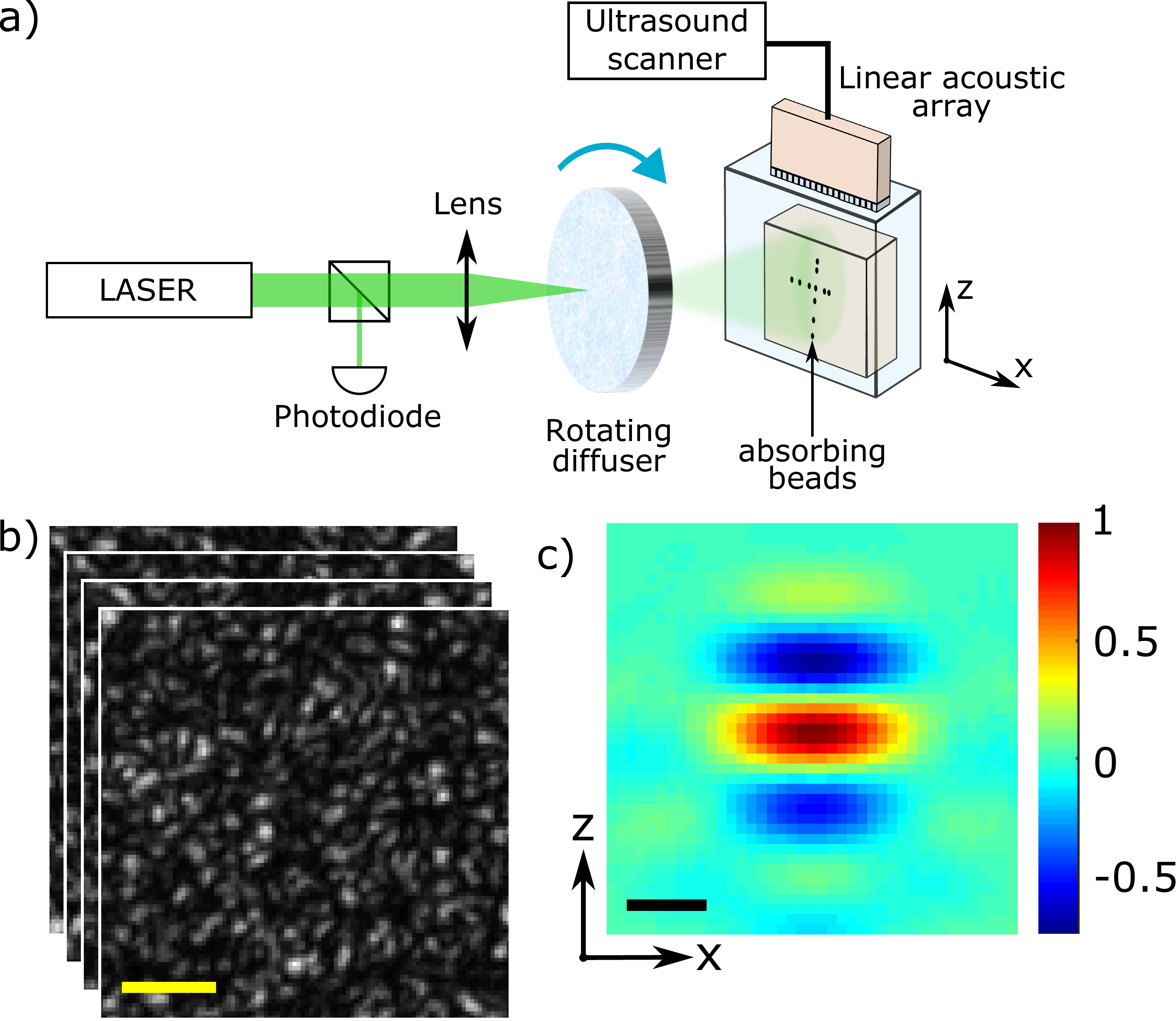}}
\caption{a) Experimental setup: A 5 ns laser pulse is focused on a rotating diffuser. The resulting speckle pattern illuminates a collection of absorbing beads, generating ultrasounds which are detected with a linear ultrasonic array. b) Typical speckle patterns (speckle grain size $\sim $30 µm). c) Typical point spread function (PSF) of the PA imaging setup (envelope FWHM: $360\pm 25$ µm in the transverse (x) direction and $450\pm 25$ µm in the axial (z) direction). Scale bars: 200 µm.}
\label{montage}
\end{figure}

Our objective was to demonstrate experimentally that the measurement of a two-dimensional (2-D) variance image can provide a super-resolution PA image of the absorption distribution beyond the acoustic-diffraction limit. The experimental setup used is shown in Fig.~\ref{montage}.a. The beam of a nanosecond pulsed laser (Continuum Surelite II-10, 532 nm wavelength, 5 ns pulse duration, 10 Hz repetition rate) was focused on a ground glass diffuser (Thorlabs, 220 grit, no significant ballistic transmission). 
The scattered light illuminated a 2-D absorbing sample embedded in an agarose gel block. This phantom was located 5 cm away from the diffuser, leading to a measured speckle grain size of $\sim$ 30 µm. The absorbing sample was placed in the imaging plane of a linear ultrasound array (Vermon, 4 MHz center frequency, >60\% bandwidth, 128 elements, 0.33 mm pitch), connected to an ultrasound scanner (Aixplorer, Supersonic Imagine, 128-channel  simultaneous acquisition at 60 MS/s). A collection of black polyethylene microspheres (Cospheric, 50 µm and 100 µm in diameter) was used to fabricate phantoms with isotropic emitters. Estimates of the PSF $h(\mathbf{r})$ were measured using isolated 50 µm diameter microspheres, while ordered patterns to be imaged were formed using 100 µm diameter microspheres. Fig.~\ref{montage}.c shows the PSF corresponding to our imaging setup. The dimensions of the PSF, defined as the full width at half maximum (FWHM) of its envelope, were $360\pm 25$ µm in the transverse (x) direction and $450\pm 25$ µm in the axial (z) direction. It is worth noting that the anisotropy and bipolarity of the PSF are singular features of limited-view and limited-bandwidth PA imaging, and do not have their counterpart in regular all-optical imaging. The measurement and analysis of the PSF $h(\mathbf{r})$ is further described in detail in Supplement 1, sec.3.

\begin{figure}
\centerline{\includegraphics[width=1\columnwidth]{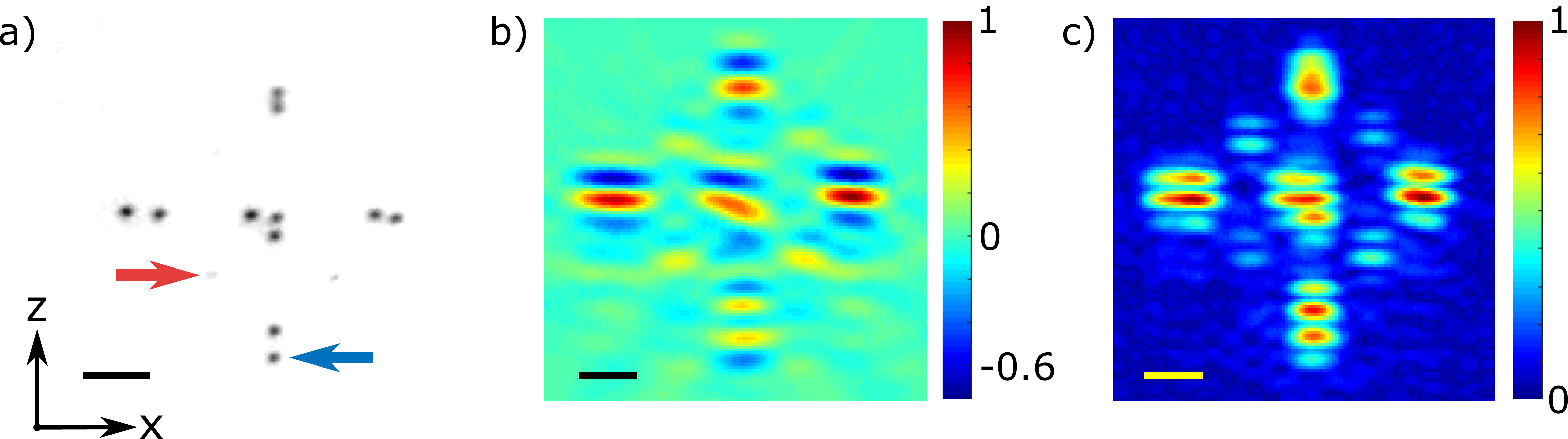}}
\caption{a) Photograph of the absorbing sample: red and blue arrows indicate respectively 50 µm and 100 µm diameter beads. b) Mean PA image over 100 speckle realisations, mimicking uniform illumination. c) Square root of the variance image over 100 speckle realisations. Scale bars: 500 µm.}
\label{fig2}
\end{figure}

For each sample, a set of PA images was reconstructed for 100 uncorrelated speckle patterns, obtained by rotating the diffuser. The mean and variance images were then computed on a per-pixel basis. As described in detail in Supplement 1, sec. 2.B, special care was taken to reduce sources of fluctuations other than the multiple speckle illumination between PA acquisitions. For the image reconstruction, a time-domain backprojection algorithm was used (see Supplement 1, sec. 2.C).  
Square images of 20 mm side were reconstructed on a grid of square pixels (25 µm side). To demonstrate the resolution enhancement of our technique, we first designed the phantom shown in Fig.~\ref{fig2}.a. Pairs of 100 µm diameter beads were positioned along the z and x axis. The distances between beads (center to center) were: 120 µm, 140 µm and 200 µm along the z direction (from top to bottom), and 250 µm, 200 µm and 160 µm along the x direction (from left to right).

Fig.~\ref{fig2}.b shows the mean PA image of the sample, obtained from averaging over the 100 speckle realizations, and which corresponds to the PA image obtained under homogenous illumination \cite{gateau2013improving}. Fig.~\ref{fig2}.c shows the square root of the variance of the same data set to make a fair comparison with the mean image in term of units and resolution. At this stage, no significant resolution enhancement can be seen from the comparison of the mean and variance images, for both of which the pairs of beads are blurred by the convolution with either $h(\mathbf{r})$ (Fig.~\ref{fig2}.b) or $h^2(\mathbf{r})$ (Fig.~\ref{fig2}.c). To demonstrate that the variance image contains sub-acoustic diffraction information thanks to the higher-frequency content of $h^2(\mathbf{r})$ as compared to $h(\mathbf{r})$, and to remove the peculiar side lobes of the PA PSF $h(\mathbf{r})$ observed in Fig.~\ref{fig2}.b and Fig.~\ref{fig2}.c, image deconvolution was performed based on Eqs.~\ref{mean} and \ref{eq:var} to retrieve the absorption distribution $\mu_a(\mathbf{r})$. The mean and variance images were deconvolved respectively by $h(\mathbf{r})$ and $h^2(\mathbf{r})$. In an ideal noise-free situation, deconvolution of an image should retrieve the absorbing object with no resolution limit. However, in the presence of noise, spatial frequencies are accurately measured up to a certain limit set by the signal-to-noise ratio (SNR). An inversion strategy that allow accounting for the presence of noise in the measurement was therefore carried out to perform the deconvolution. Retrieving $\mu_{a} ^2(\mathbf{r})$ from measurements of the variance image modeled as $  \widehat{\sigma^2[A]}(\mathbf{r}) = h^2(\mathbf{r}) \ast \mu_{a}^2(\mathbf{r}) + \varepsilon$ (with $\varepsilon$
accouting for the experimental noise) was carried out by the minimization of the following constrained least-square functional:
\begin{equation}
\label{Pen_criterion}
 J(x) := || h^2 \ast x - \widehat{\sigma^2[A]}||^2 + \alpha ||x||^2 \quad \text{subject to} \quad x \geq 0 
\end{equation}
with $||\cdot ||$ the Euclidian norm over the image space, and $\alpha\geq 0$ a regularization parameter. The constrained minimizer  $\widehat{x}_\alpha$ provides a regularized solution to the deconvolution problem, which in turn defines an estimation of the absorption distribution 
$ \widehat{\mu_a}_\alpha := \sqrt{\widehat{x}_\alpha}$. For comparison, the exact same approach was applied to retrieve an estimation of $\mu_a$ from the measurement of the mean image. In practice, the minimization of the functional above required adjusting the regularization parameter $\alpha$ in order to obtain a fair resolution \textit{v.s.} noise trade-off~\cite[Sec. 5.6]{Bertero98}, and choosing $N_{\mathrm{iterations}}$ in the FISTA numerical method used to perform the minimization (see Supplement 1, sec. 1.C for further details on the minimization approach and the corresponding algorithm). 

\begin{figure}
\centerline{\includegraphics[width=1\columnwidth]{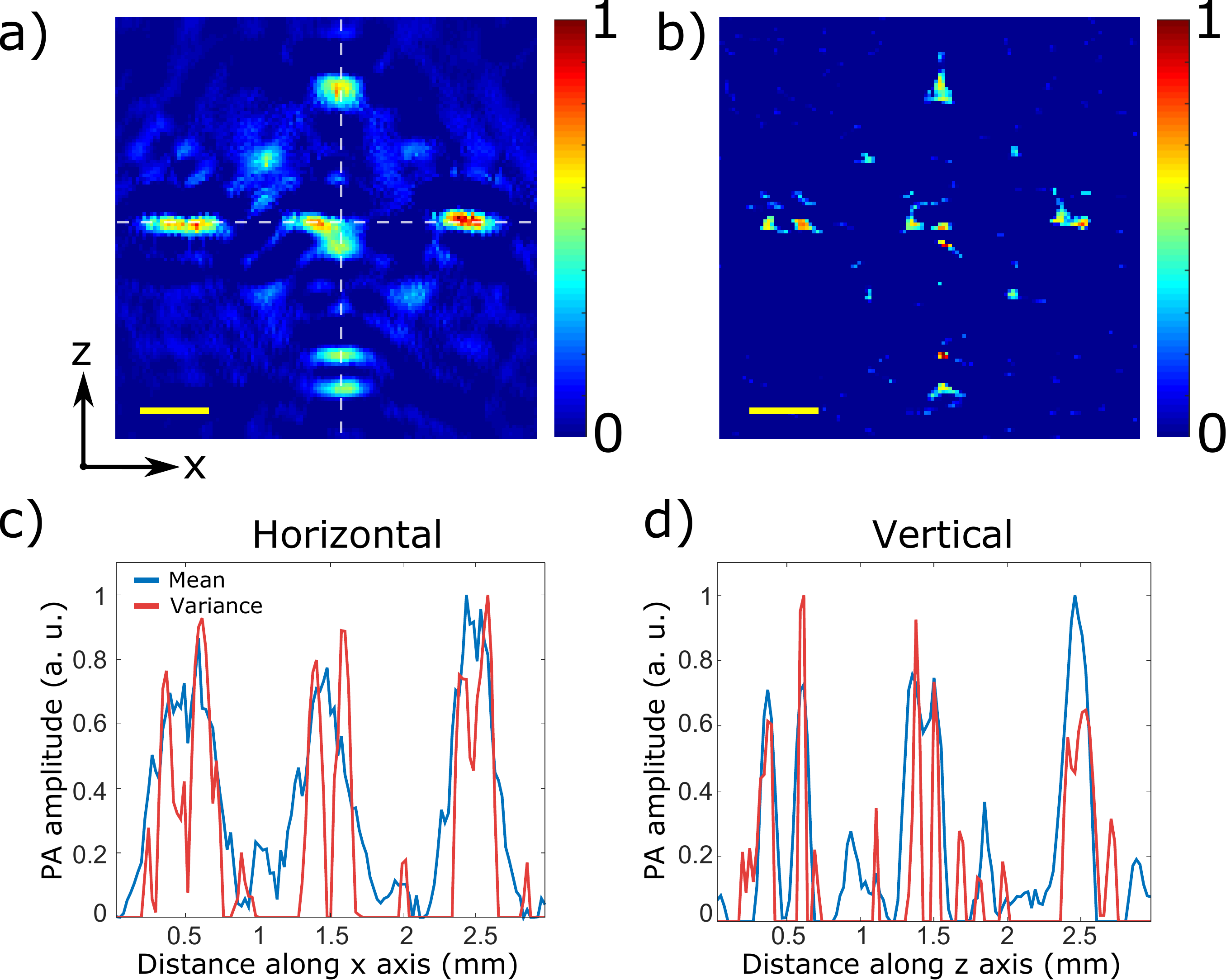}}
\caption{a) Mean image deconvolved by the PSF; white dashed lines indicates the cross-section directions. b) Square root of the variance image deconvolved by the squared PSF. c) Horizontal cross-sections, blue curve: deconvolved mean image, red curve: square root of deconvolved variance image. d) Vertical cross-sections, blue curve: deconvolved mean image, red curve: square root of deconvolved variance image.  Scale bars: 500 µm.}
\label{fig3}
\end{figure}

Deconvolved images are shown in Fig.~\ref{fig3}. As mentioned above, the deconvolved variance image is square rooted to retrieve the absorption distribution $\widehat{\mu_a}_\alpha $. The parameters of the deconvolution algorithm were $\alpha =10^5,\  N_{\mathrm{iterations}}=800$. On the deconvolved mean image (Fig.~\ref{fig3}.a), we observe that only absorbers separated by the largest distance in each direction are resolved (z direction: 200 µm, x direction: 250 µm). In comparison, the absorbers appear much sharper on the deconvolved variance image (Fig.~\ref{fig3}.b) and almost every pair is resolved on this image. The only exception is the upper one, for which the beads are nearly touching. Cross-sections along both directions are shown respectively on Fig.~\ref{fig3}.c and d.
These cross-sections show the maximum amplitude projection of the PA image along a $\pm 100 \mu m$ band around the white dashed lines plotted in Fig.~\ref{fig3}.a.

\begin{figure}
\centerline{\includegraphics[width=1\columnwidth]{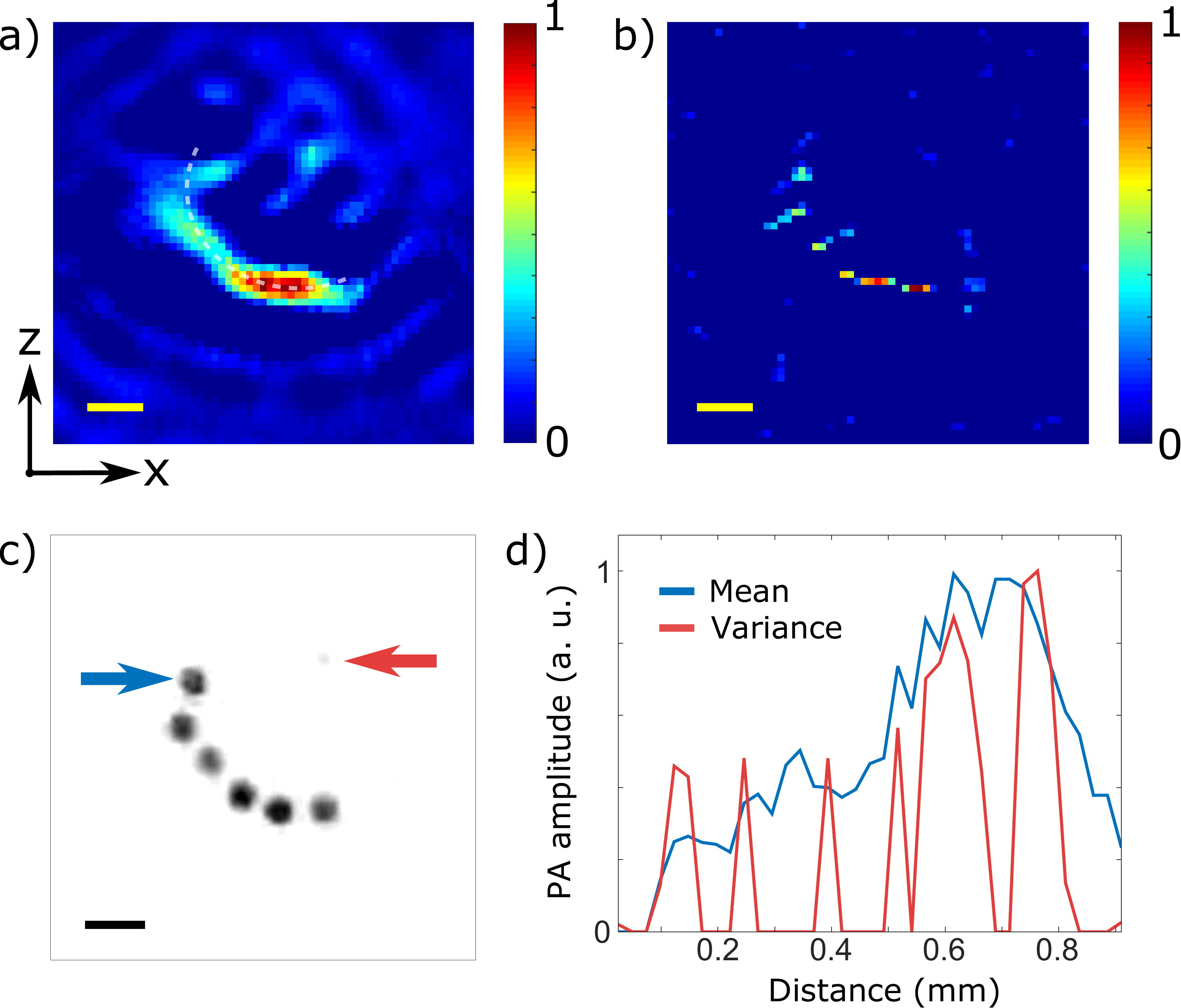}}
\caption{a) Mean image deconvolved by the PSF. b) Square root of the variance image de convolved by the squared PSF; white dashed lines indicates the cross-section direction. c) photograph of the absorbing sample, red and blue arrows indicate respectively 50 µm and 100 µm diameter beads. d) Cross-sections, blue curve: mean image, red curve: square root of variance image. Scale bars: 200 µm.}
\label{fig4}
\end{figure}

To further investigate the influence of the PSF anisotropy, a second phantom was designed, in which 100 µm diameter beads are equally spaced from each other with a center-to-center distance of about 150 µm and distributed on a quarter circle. The corresponding results are shown in Fig.~\ref{fig4}.  The parameters of the deconvolution algorithm were $\alpha =10^6,\ N_{\mathrm{iterations}}=600$. Not a single bead can be resolved on the deconvolved mean image (Fig.~\ref{fig4}.a). In contrast, the 6 beads can easily be distinguished on the deconvolved variance image (Fig.~\ref{fig4}.b). The last two beads on the right are still distinguishable but tend to smear, according to the lower resolution in the transverse x direction. Because of the PSF anisotropy (a consequence of the limited view), we observe that the resolution strongly depends on the angle of the beads pair with respect to the probe surface.

Speckle-induced PA fluctuation imaging is an original method recently proposed by our group to enhance visibility in PA imaging. In this study we demonstrated that fluctuation imaging can also be used to overcome the ultrasound diffraction limit in PA tomography. 
We showed that this method can separate small absorbers unresolved under standard uniform illumination. This is to the best of our knowledge the first super-resolution PA imaging method which does not require optical focusing, and therefore is applicable beyond the ballistic regime.
Super resolution PA microscopy was reported in previous studies, but it relied on light focusing and on nonlinear absorption mechanisms, a situation which cannot be translated to deep tissue PA imaging \cite{rao2011real,wang2014grueneisen}.

The common approach to obtain high resolution at depth in PA imaging was to date to improve the acoustic detection by employing higher frequencies and increasing the detection aperture. Here, we demonstrated that additional efforts can be deployed towards the illumination, in order to induce fluctuations and separate discrete absorbers below the acoustic diffraction limit.

Deconvolution was found essential to reveal the super-resolution properties with the second-order fluctuation images.
Deconvolution has already been considered in PA imaging, as part of the reconstruction algorithm \cite{zhang2010fast} and to compensate for smearing induced by the spatial impulse response of the finite-size detectors \cite{roitner2014deblurring}.  
However, no super-resolution was yet demonstrated , since there were no physical mechanism extending the high spatial frequency information.
Our approach goes beyond past works by considering fluctuations in PA images as a signal that reveal higher spatial frequencies above the noise level. A gain in resolution was obtained in both transverse and axial directions, with a resolution enhancement of at least 1.7 in both directions.

The deconvolution approach was implemented with absorbers of size similar to that of the absorbers used to measure the point spread function. As an immediate drawback, deconvolution was unable to restore the actual size of the absorbers, which lead on the deconvolved image to reconstructed beads smaller than their real size.  Nonetheless, this method showed a very good sub-acoustic resolution performance, which was the objective of our work. Although the beads were expected to be almost equally absorbing, a difference in the reconstructed amplitudes was noticed.  The proposed method was therefore not shown to be quantitative, which could be attributed to the non-linear deconvolution scheme. Further work should be carried out to investigate the possibility to retrieve quantitative absorption information.

Our proof-of-concept experiments were carried out at low medical ultrasound frequency to simplify the controlled fabrication of absorbing samples (see Supplement 1, sec. 2.A). However, the approach could be scaled down using high frequency detectors, and could also be extended to 3D PA imaging.
It must be emphasized that speckle illumination is achievable at depth using a seeded laser \cite{bjorkholm1969frequency}, whose long coherence length still provides speckle formation after several centimeters of propagation in scattering tissue.
PA fluctuations induced by speckle illumination are expected to decrease as the number of speckle grains contained in the absorbers increases~\citep{gateau2013improving}. Detection of speckle-induced fluctuations would therefore be challenging deep inside tissue where the speckle grains are on the order of half the optical wavelength, and will require special care on the excitation and detection stability.
In the past few years, several techniques combining shaped coherent illumination and PA imaging have been developed. These methods aim at focusing light inside tissue, possibly with sub-acoustic resolution, which could also lead to super-resolved PA imaging \cite{chaigne2014controlling,caravaca2013high,lai2015photoacoustically}. However, these methods require expensive hardware and tedious experimental procedures. We proposed here a very simple imaging technique than does not require any costly equipment and nearly no optical alignment. For biological applications, tissue-induced temporal decorrelation of speckle patterns could even be exploited as a source of fluctuating illumination \cite{jang2015relation}. Alternatively, this super-resolution method can be extended to fluctuation of the absorption induced by blinking or switchable~\cite{ng2014stimuli} contrast agents.

\paragraph{Acknowledgements.} This work was funded by the European Research Council (grant 278025), the Fondation Pierre-Gilles de Gennes (grant FPGG031) and by the LABEX WIFI (Laboratory of Excellence ANR-10-LABX-24) within the French Program “Investments for the Future” under reference ANR-10- IDEX-0001-02 PSL*. O.K. acknowledges the support of the Marie Curie Intra-European Fellowship. The authors thank Laurent Bourdieu and Jean-François Léger for their helpful assistance on the milling machine.\\

\paragraph{Supplementary materials.} See Supplement 1 for supporting content.
\bibliography{citation_database}

\clearpage
\onecolumn
\begin{center}
\textbf{\Huge Super-resolution photoacoustic fluctuation imaging with multiple speckle illumination: supplementary material}
\end{center}
\setcounter{equation}{0}
\setcounter{figure}{0}
\setcounter{table}{0}
\makeatletter
\renewcommand{\theequation}{S\arabic{equation}}
\renewcommand{\thefigure}{S\arabic{figure}}
\renewcommand{\bibnumfmt}[1]{[S#1]}
\renewcommand{\citenumfont}[1]{S#1}

\section*{}
This document provides supplementary information to “Super-resolution photoacoustic fluctuation imaging with multiple speckle illumination”. The first section contains theoretical informations: the conditions are given to model the photoacoustic images reconstruction as a convolution, the variance image is shown to result from the convolution of the squared PSF, and details on the deconvolution algorithm used are given. The second section provides more detailed information on the experimental methods, including samples preparation, measurements and processing of photoacoustic signals, and the backprojection algorithm used for the photoacoustic reconstruction. The last section analyze some properties of the photoacoustic point spread function.

\section{Theory}
\label{Theory}

\subsection{Photoacoustic imaging as a convolution process}
\label{PA as a convolution}

In its simplest form, the generation and propagation of photoacoustic (PA) pressure waves $p(\mathbf{r},t)$,  generated by a distribution $\mu_{a}(\mathbf{r})$ of optical absorption illuminated by a pulsed light with intensity $I(\mathbf{r})\times f(t)$, may be described in a homogenous acoustic medium (with speed of sound $c_s$ and Grüneisen coefficient $\Gamma$) by the following equation
\begin{equation}
\label{eq:PAeq}
\left[\frac{\partial^2}{\partial t^2}- c_s^2\nabla^2\right] p(\mathbf{r},t) = \Gamma \mu_a(\textbf{r}) I(\mathbf{r}) \frac{\partial f(t)}{\partial t}
\end{equation} 
Eq.\ref{eq:PAeq} shows that for a given temporal pulse shape $f(t)$, the pressure field is \textit{linearly} related to the distribution $\mu_{a}(\mathbf{r})\times I(\mathbf{r})$ of optical absorption times the optical intensity pattern. Various algorithms exist that aim at reconstructing $\mu_{a}(\mathbf{r})\times I(\mathbf{r})$ from a set of PA signals measured at various points located on some boundary around the sample to be imaged~\cite{Xu2006Photoacoustic,rosenthal2013acoustic}. In the context of our work, we consider a conventional back-projection algorithm based on summing values of the PA signals taken at appropriate retarded times~\cite{Xu2006Photoacoustic}. As such, the reconstructed images remain \textit{linearly} related to $\mu_{a}(\mathbf{r})\times I(\mathbf{r})$. Moreover, if this linear reconstruction process may also be considered as translation-invariant, i.e. a spatial translation of the object simply translates the reconstructed object, then the reconstruction process may be written as a convolution as assumed by Eq. 1 of our work. Because the PA signals are measured on some boundary around the region to be imaged, the translation invariance cannot be strictly verified. However, for a field of view small enough such that every point that it contains is reconstructed with the same point spread function (PSF), it  becomes possible to reasonably assume that the reconstruction process is translation-invariant. In our case, this assumption was validated by measuring PSF at four different locations in the field of view. The four PSF appeared identical (see sec.~\ref{Invariance of the PSF} below), and the results shown in the manuscript were independant of the particular PSF that was used to perfom the deconvolutions, thus validating our assumption that the reconstruction process may be written as a convolution.

\subsection{The variance image as a convolution with $h^2(\mathbf{r})$}
\label{Variance demo}

From $A(\mathbf{r})=\left[\mu_{a}(\mathbf{r})\times I(\mathbf{r})\right]\ast h(\mathbf{r})
$ (valid only for objects lying in a small enough field of view, see sec.~\ref{PA as a convolution} above), the variance image $\sigma^2[A](\mathbf{r})$ reads
$$
\sigma^2[A](\mathbf{r})=\iint C(\textbf{r}',\textbf{r}'') \mu_{a} (\textbf{r}') \mu_{a} (\textbf{r}'') h(\textbf{r}-\textbf{r}') h(\textbf{r}-\textbf{r}'') d\textbf{r}' d\textbf{r}''
$$
where 
$$C(\textbf{r}',\textbf{r}'')= \langle I(\textbf{r}')\cdot I(\textbf{r}'') \rangle - \langle I(\textbf{r}') \rangle \cdot \langle I(\textbf{r}'') \rangle $$ 
is the autocovariance of the intensity fluctuation in the speckle patterns. Under the common assumption that the speckle is wide-sense stationary \cite[p.67]{goodman2007speckle}, the autocovariance function $C$ is a function of only one variable $\textbf{r}=\textbf{r}'-\textbf{r}''$. This function $C(\textbf{r})$ is sharply peaked around its center, and its characteristic length is by definition the speckle grain size \cite[p.72]{goodman2007speckle}. If the speckle grains are small enough compared to the PSF, $C(\textbf{r}'-\textbf{r}'')$ may be considered in the double integral above proportional to a delta function $\delta(\textbf{r}'-\textbf{r}'')$, yielding the following convolution expression:

$$
\sigma^2[A](\mathbf{r})\propto\int \mu_{a}^2 (\textbf{r}')h^2(\textbf{r}-\textbf{r}') d\textbf{r}' =\mu_{a} ^2(\mathbf{r}) \ast h^2(\mathbf{r})
$$
This expression is strictly equivalent to that found for the second-order analysis in SOFI~\cite{dertinger2009fast}, apart from the origin of the fluctuations.

\subsection{Deconvolution algorithm: FISTA}
\label{FISTA}

As introduced in the main manuscript, deconvolution of the variance image was performed  by minimizing the following constrained least-square functional:
\begin{equation}
\label{Pen_criterion}
 J(x) := || h^2 \ast x - \widehat{\sigma^2[A]}||^2 + \alpha ||x||^2 \quad \text{subject to} \quad x \geq 0 
\end{equation}
From a practical viewpoint, an iterative minimization algorithm is required for the numerical evaluation of $\widehat{x}_\alpha$. Since $J$ is a strictly convex functional if $\alpha>0$,  its global minimizer $\widehat{x}_\alpha$ is asymptotically reached by any locally convergent iteration, whatever the initial-guess of the algorithm. For this constrained optimization task, a natural candidate is the standard projected-gradient (PG)  since its computational burden is very low. The PG is however rather slow to converge and the FISTA iteration~\cite{Beck09} that achieves a faster convergence is presently considered as a popular alternative and was used in our work. In comparison to the standard projected-gradient algorithm~\cite{Bertsekas99}, the FISTA method~\cite{Beck09} only requires the additional storage of an auxiliary component
$y^{(k)}$. Let $x^{(0)}$ be an initial-guess and $y^{(0)}=x^{(0)}$,
the FISTA update for $k=0,1,\cdots$ is: 
\begin{equation}
  \label{algo}
  \left \{
  \begin{array}{rcl}
    x^{(k+1)} &=& \mathcal{P}_+\left( y^{(k)} - \theta g(y^{(k)}) \right)\\[.5em]
    y^{(k+1)} &=& x^{(k+1)} + \frac{k-1}{k+2} \left(x^{(k+1)} - x^{(k)} \right)
  \end{array}
  \right.
\end{equation}
with $\theta>0$, $\mathcal{P}_+$   the projection operator over $\mathbb{R}_+$,
and $g$ the gradient of the penalized least-square functional above that reads
\begin{equation}
  \label{grad}
  g(x) \, := \,   2 \left(h^2\right)^\dag \ast \left( h^2 \ast x - \widehat{v}\right) + \alpha x 
\end{equation}
where  $h^\dag$ stands for the adjoint of the operator $h$. It is worth noting that all the convolution operations that appear above can 
be computed efficiently in the Fourier domain \textit{via} the fast Fourier transform (FFT) 
and appropriate boundary conditions \cite{Bertero05}. Finally, the global convergence of 
\eqref{algo} is granted provided that the constant step-size $\theta$ is adjusted within 
the interval $(0,\overline{\theta})$ with 
$$\overline{\theta}\equiv 1/\max_\omega(|\widetilde{h}(\omega)|^2 + \mu)$$  
where $\widetilde{h}(\omega)$ stands for the Fourier transform of $h$.

\section{Experimental methods}
\label{Experimental methods}

\subsection{Samples preparation}
\label{Samples}

A collection of black polyethylene microspheres (Cospheric, 50 µm and 100 µm in diameter) was used to fabricate phantoms with isotropic emitters. Estimates of the PSF $h(\mathbf{r})$ were measured using 50 µm diameter microspheres, while ordered patterns to be imaged were formed using 100 µm diameter microspheres. To precisely control the relative position of the beads, melted gel was first poured on a mold drilled with micrometer precision (Mini Mill, Minitech). The beads were then manually placed in the molded holes of the solidified gel.

\subsection{Measurements and signal processing}
\label{Measurements and signal processing}

To avoid other sources of fluctuation apart from the multiple speckle illumination, appropriate care was taken to eliminate noise and triggering-induced temporal jitters. The intensity of the laser pulses was monitored with a photodiode to compensate for the laser pulse-to-pulse energy fluctuations. In addition, for each speckle illumination (i.e. each diffuser position), the PA signals were averaged over 25 laser pulses to improve the signal-to-noise ratio. The signals were then filtered between 1 MHz and 8 MHz ($3^{rd}$ order butterworth filter) for noise removal outside of the array bandwidth. A set of PA images was reconstructed for 100 uncorrelated speckle patterns. The mean and variance images of this data set were then computed on a per-pixel basis. The same procedure was repeated with one single speckle realization (static diffuser). The resulting variance image was then subtracted from the previous variance image (with rotating diffuser). This procedure was found to correct for systematic variance noise.

\subsection{Backprojection algorithm}
\label{backprojection}

In the context of our work, we considered a conventional back-projection algorithm based on summing values of the PA signals taken at appropriate retarded times. More specifically, the backprojection algorithm described  in \cite[eq. (20)]{xu2005universal} was implemented while keeping only the first time-derivative of the processed signal, assuming point-like detectors located in the centers of the elements of the ultrasound array.

\section{Photoacoustic point spread function}
\label{PSF}

\subsection{Measurement}
\label{PSF measurement}

The PSF of the imaging setup was measured by concentrating light on a single 50 µm diameter bead located in the vicinity of the structured 100 µm bead pattern (see Fig.~2.a). This ensured that the 50 µm microsphere was the only PA source. The diffuser was removed from the light path during this step. 

\subsection{Translation-invariance of the point spread function across the field-of-view}
\label{Invariance of the PSF}

\begin{figure}[h!]
\centerline{\includegraphics[width=0.7\columnwidth]{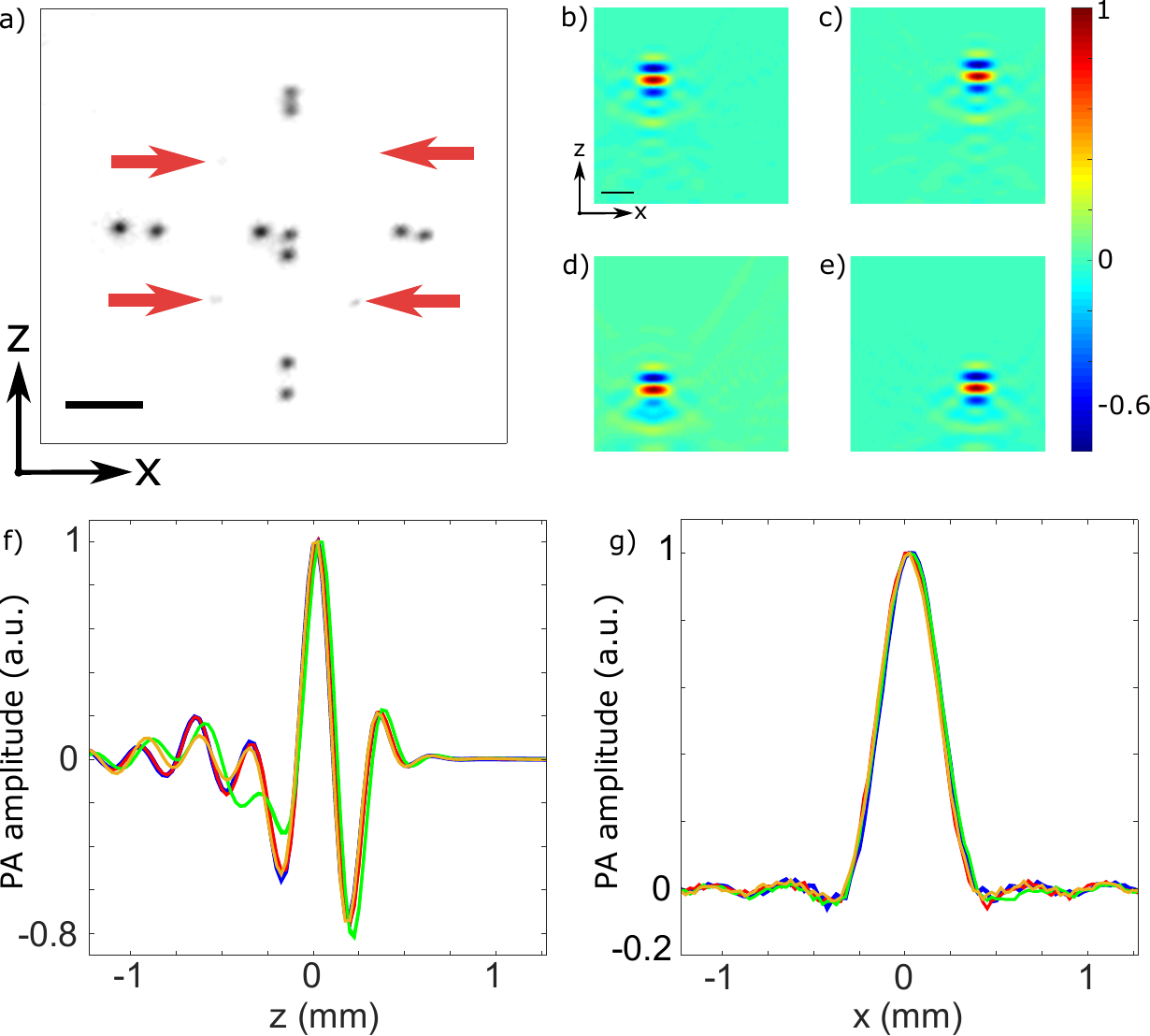}}
\caption{a) Photograph of the absorbing sample. The red arrows indicate the locations of the 50 µm diameter beads used to measure the PSF of the PA system. b-e) PSF of the PA imaging setup, recorded at 4 different locations: (b) Top left, (c) Top right, (d) Bottom left, (e) Bottom right. f) Axial cross-sections of the 4 different PSFs (along z direction): top-left (blue), top-right (red), bottom-left (green), bottom-right (yellow). g) Transverse cross-sections of the 4 different PSFs (along x direction). Scale bars: 500 µm.}
\label{figureS3}
\end{figure}

Following the measurement approach described above, PSFs were reconstructed for 4 different locations in the field of view to confirm that it can be assumed as constant, as required to model the reconstruction by a convolution process. The corresponding results are summarized in Fig.~\ref{figureS3}. With our current ultrasonic detection and the back-projection algorithm, we showed that this approach was at least valid for a 3 mm x 3 mm field-of-view. On the deconvolved images, a slight difference in the PSF shapes would results in additional artefacts, for instance some rebounds around the reconstructed objects that are located far from the place where the PSF is measured.

\subsection{Analysis}
\label{PSF analysis}

\begin{figure}[h!]
\centerline{\includegraphics[width=0.8\columnwidth]{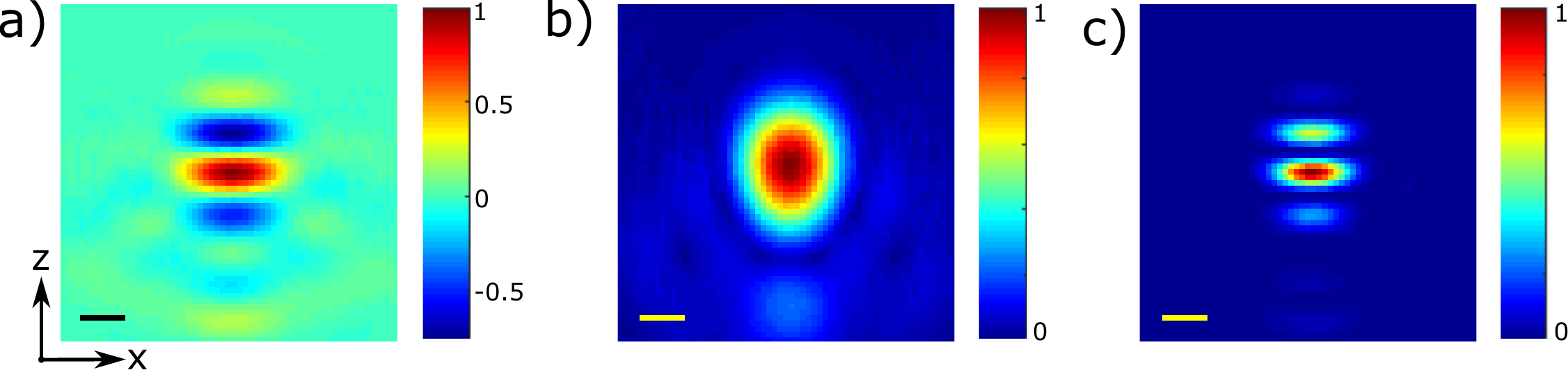}}
\caption{a) PSF of the PA imaging system. b) Enveloppe of the PSF (modulus of the Hilbert transform). c) Squared PSF of the PA imaging system. Scale bars: 200 µm.}
\label{figSpsf}
\end{figure}

\begin{figure}[h!]
\centerline{\includegraphics[width=0.7\columnwidth]{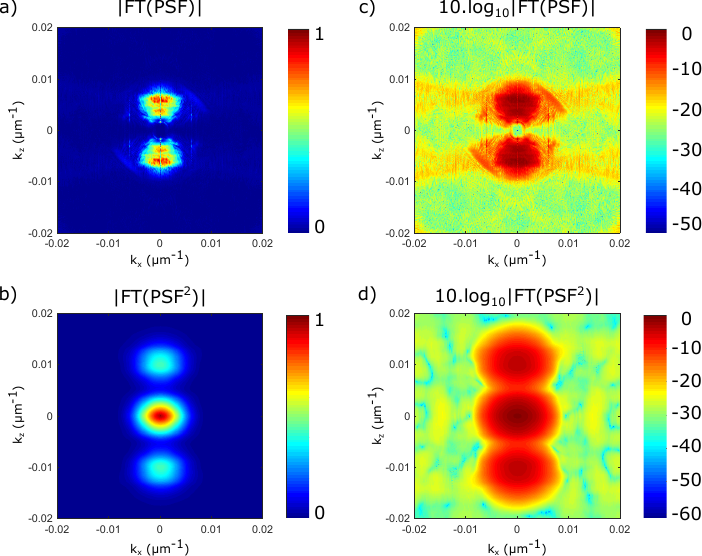}}
\caption{a) Two-dimensional spatial Fourier Transform (magnitude) of the PSF. b) Two-dimensional spatial Fourier Transform (magnitude) of the squared PSF. c-d) Same as a-b) with logarithmic scale.}
\label{figSspec}
\end{figure}

In Fig.~\ref{figSpsf}, we display the PSF (Fig.~\ref{figSpsf}.a) of the PA system and its envelope (Fig.~\ref{figSpsf}.b, computed using Hilbert transform), and the squared PSF (Fig.~\ref{figSpsf}.c). The conventional resolution of the imaging system (when using uniform illumination) is given by the full width at half maximum (FWHM) of the enveloppe of the PSF.  The FWHM measured on the data shown on Fig.~\ref{figSpsf}.b was $360\pm 25$ µm in the transverse x direction and $450\pm 25$ µm in the axial z direction. On Fig.~\ref{figSpsf}.c, we observe that the squared PSF is sharper than the PSF itself, which is expected to lead to a higher measurable frequency content in the fluctuation image. The spatial frequency content of the PSF and the squared PSF of the PA imaging setup are shown in Fig.~\ref{figSspec}. We observe that for a given noise level, higher spatial frequencies are measurable with the squared PSF, hence on the variance image.

\vspace{50pt}

\bibliography{citation_database}

\end{document}